\documentclass[3p,review]{elsarticle} 

\usepackage{amsmath}
\usepackage{pgfplots}
\usepackage{subcaption}
\usepackage{textcomp}
\usepackage{tikz}
\usepackage{todonotes}
\usepackage[colorlinks,bookmarksopen,bookmarksnumbered,citecolor=red,
urlcolor=red,breaklinks=true]{hyperref}

\tikzset{>=latex}

\begin{document}

\begin{frontmatter}

\title{Pixel-Level Statistical Analyses of Prescribed Fire Spread}
\author[gfdi]{M.~Currie}
\author[gfdi,eaos]{K.~Speer}
\author[ttrs]{J.~K.~Hiers}
\author[usfs]{J.J.~O'Brien}
\author[usfs]{S.~Goodrick}
\author[gfdi,dsc]{B.~Quaife\corref{cor}}\ead{bquaife@fsu.edu}

\address[gfdi]{Geophysical Fluid Dynamics Institute, Florida State University, Tallahassee, FL}
\address[eaos]{Earth, Ocean and Atmospheric Sciences, Florida State University, Tallahassee, FL}
\address[ttrs]{Tall Timbers Research Station, Tallahassee, FL}
\address[usfs]{U.S.~Forest Service, Southern Research Station, Athens,
GA}
\address[dsc]{Department of Scientific Computing, Florida State
University, 400 Dirac Science Library, Tallahassee, FL, 32306}
\cortext[cor]{Corresponding author.}

\begin{abstract}
  Wildland fire dynamics is a complex turbulent dimensional process.
  Cellular automata (CA) is an efficient tool to predict fire dynamics,
  but the main parameters of the method are challenging to estimate.  To
  overcome this challenge, we compute statistical distributions of the
  key parameters of a CA model using infrared images from controlled
  burns.  Moreover, we apply this analysis to different spatial scales
  and compare the experimental results to a simple statistical model.
  By performing this analysis and making this comparison, several
  capabilities and limitations of CA are revealed.
\end{abstract}


\end{frontmatter}

\section{Introduction}
Wildland fire dynamics involve both complex fuel geometries and
three-dimensional turbulent flow. Measuring, modeling, and predicting
the behavior of fire under realistic conditions is a major technical and
computational challenge. The fundamental impact of turbulence in the
surface boundary layer and in the self-generated convective flow leads
to difficulties relating the smaller scale flow and heat transfer to
larger scale fire spread.  Understanding these dynamics is nevertheless
essential for natural resource conservation, and for developing tools to
protect life and property.

Much effort has been put into developing efficient numerical simulations
on both discrete systems, such as cellular automata
(CA)~\cite{wolfram1983, ach2003, albinet1986, duarte1997, sul-kni2004},
and continuous systems, typically represented by a set of partial
differential equations (PDEs)~\cite{watt1995, richards1990, mar-ser1998,
lin-rei-col-win2002}, or using both in conjunction \cite{men-lle1997,
clarke1994}. PDE models incorporate the complex physics involved in
wildfire, such as the chemistry, radiation, and fluid dynamics, but
these methods can be too computationally expensive to be used in a
real-time framework.  Alternatively, the discrete, lattice-type
approach, with individual elements or cells having defined states (for
example unburnt and burnt) has the advantage of being computationally
fast, but may be difficult to relate to the actual properties of the
fuel and other environmental conditions.

The basic CA approach, with probabilities for changing state defined by
some rules, is the simplest case of a discretized model governing the
behavior of fire propagation~\cite{sul2009}.  The possible states of a
cell in a CA model for wildfires includes burnt, ignited, and
combustible.  Two of the most important parameters for the basic CA
model are the transition probabilities between these states, and the
amount of time a cell remains in each of the states (burn time).  We
compute spatial statistics of probabilities by analyzing infrared (IR)
images from prescribed burns.  The images were derived from an IR camera
with a spatial resolution of 1cm$^{2}$ and a time resolution of 1 Hz.
These images have been successfully used for other studies such as
O'Brien et al.~\cite{obr-etal2016} who analyzed the radiant power
output.

\paragraph{Related Work}
CA codes have been used in the past to understand fire dynamics with the
commonality that each of these models have to choose rules for the
propagation of the fire.  The main choices are the probability that a
fire spreads from a burnt to an unburnt cell, which cells the fire can
spread to, the amount of time that a burning cell remains burning, and a
fire-atmosphere coupling.  While it is clear that the probabilities of
spread are important, the burn time also plays a significant role, and
brings with it a measure of fuel load.  With increasing burn time comes
increasing probability of the combustion of a neighboring area.  In
addition, the spatial scales: resolution and domain size, and effective
time step of the CA model must be chosen.

Achtemeier~\cite{ach2003,ach2013} developed a CA model called ``Rabbit
Rules''.  The model draws a parallel between a  propagating fire and a
population of rabbits that are moving, reproducing, and dying.  In
addition, nonlocal effects are included by coupling the heated plume
generated by the fire to either a model for the
atmosphere~\cite{ach2005} or meteorological measurements.  Dahl et
al.~\cite{dah-xue-hu-xue2015} developed a coupled fire-atmosphere
approach using CA called a raster-based methodology.  Their spread rate
is computed from a number of input variables using the Rothermel
model~\cite{rot1972}.  Both ``Rabbit Rules" and raster-based models
represent fire dynamics as a complex diffusive process, and the
characteristics of this diffusion depend on the nature of the
probabilities used for driving the ignition and combustion of natural
fuels, the burn times of these fuels, and the probability of new
ignition sites through spotting.  Almeida and Macau~\cite{alm-mac2011}
presented a stochastic CA model that contains several parameters thought
to be relevant to the fire spread.  Their ``D" is related to fuel
concentration, ``I" is probability of ignition, and ``B" is the
probability that a cell remains burning once ignited.  They make the
argument that this persistence is also a combustion latency, or
potential to ignite a neighbor, in turn a function of the burn time.

Trunfio et al.~\cite{tru-dam-ron-spa-dig2011} showed that it is possible
and overall more advantageous to utilize a cell-based fire spread model
that effectively does not limit the size or shape of a single cell.
This mitigates the constrained directions of travel and neighborhood
size given by the inherent shapes and angles of a cell which is constant
in shape and size. With less computational cost, a grid of dynamic cell
shapes is a significant improvement over methods with static cell
shapes.

When combined with PDE models, CA can be an extremely powerful tool for
investigating fire spread.  An example of this is diffusion-limited
aggregation models, which Clarke et al.~\cite{clarke1994} compared to
real experimental burns and found that the pixels in the simulation
capture the correct result roughly 80\% of the time. More recently,
groups have been actively pursuing more data-driven CA models which
incorporate real fire data to improve accuracy~\cite{ntinas17,
mahmoud17}.

Despite these advances, large numbers of parameters remain chosen by
some combination of empirical evidence or objective fitting procedure,
typically based on on integral measures such as burn area rather than
the detailed transitions from one cell to another.  In this paper, we
use robust experimental data from relatively small burn plots with light
to moderate fuel characteristics to determine the detailed statistical
properties of the fire spread under various conditions.  The analysis
provides several essential parameters that are required in a CA code.
In addition, we analyze the validity of local versus nonlocal transition
probabilities at several different scales for these particular
prescribed burns.  By computing the parameters for an empirical CA
method, the method can be potentially applied to conditions outside the
range of the data~\cite{cru-ale-sul2017}.

\paragraph{Outline of the paper}
In Section~\ref{sec:stats}, parameters of a CA method are extracted from
experimental data.  Section~\ref{sec:simpleModel} describes a simple
model for predicting fire spread and it is compared with experimental
results.  Section~\ref{sec:aggregation} analyzes the effect of scale by
considering the CA parameters of aggregated cells.  Finally, the results
are further discussed and concluding remarks are made in
Section~\ref{sec:discussion}.

\section{Extracting Cellular Automata Parameters}
\label{sec:stats}
We present several basic statistics that are extracted at the pixel
level from an IR time series of a controlled burn. Namely, we determine
the probabilities of a pixel igniting (assuming the pixel has the
potential of lighting on fire, i.e.~necessary fuel) when its closest
neighbors are burning in the previous time step, thus quantifying the
probability and direction of wildfire spread. We also compute burn time
distributions of the pixels.

In all, 40 datasets comprising of time series of IR temperature of
experimental prescribed burns are analyzed. Hence, each dataset is a
time series of IR images where each pixel is a temperature value at a
given 1cm $\times$ 1cm location in space and time.  For each dataset, we
construct distributions for the burn time, and the direction and
probability of spread.  In Section~\ref{sec:simpleModel}, we compare the
experimental data to a simple local model that captures the aggregated
fire spread in these experiments.  Finally, in
Section~\ref{sec:aggregation}, we examine any changes in these
distributions when pixels are aggregated to form larger pixels.

\subsection{Basic Statistical Analysis}
\begin{figure}
\centering
\includegraphics[width=0.8\textwidth]{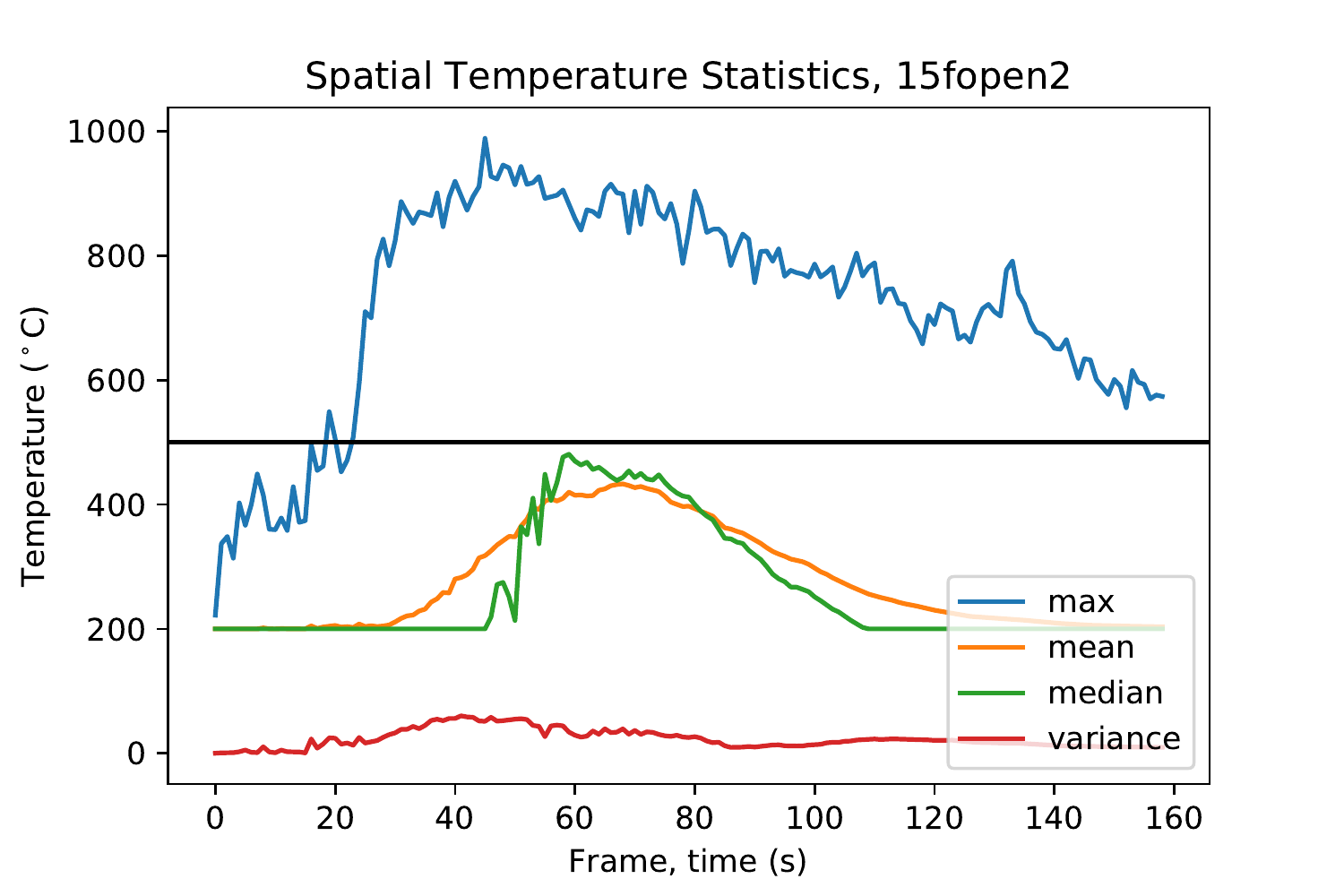}
\caption{\label{fig:basicStats} Basic temperature statistics of a single
dataset ({\em 15fopen2}). The IR camera is calibrated to measure
temperatures above 200\textdegree{}{C}.  The black line indicates a
temperature of 500\textdegree{}{C} that we use to distinguish between
burning and non-burning pixels.}
\end{figure}

We start by computing basic spatial statistics of a time series of
images: the mean, median, maximum, and standard deviation of the
temperature values of each frame. To illustrate the typical behavior, in
Figure~\ref{fig:basicStats} we plot these quantities for the dataset
{\em 15fopen2}, a prescribed fire from 2015 in a longleaf pine flatwoods
forest plot with no trees in the field of view. As the IR camera was
calibrated to measure temperatures above 200\textdegree{}{C}, the
results are determined with this background rather than ambient
temperature. For this particular fire, there is a sudden increase in the
mean and maximum temperature near $t=20$s which indicates that a
significant number of pixels have ignited. We use the maximum
temperature achieved at this time, about 500\textdegree{}{C}, to define
a threshold between the burning and non-burning states. The choice is
somewhat arbitrary in the sense that heterogeneous fuels imply a range
of ignition temperatures; other choices are possible but lead to similar
results.

\subsection{Pixel Burn Times}
\begin{figure}
\centering
\includegraphics[width=0.8\textwidth]{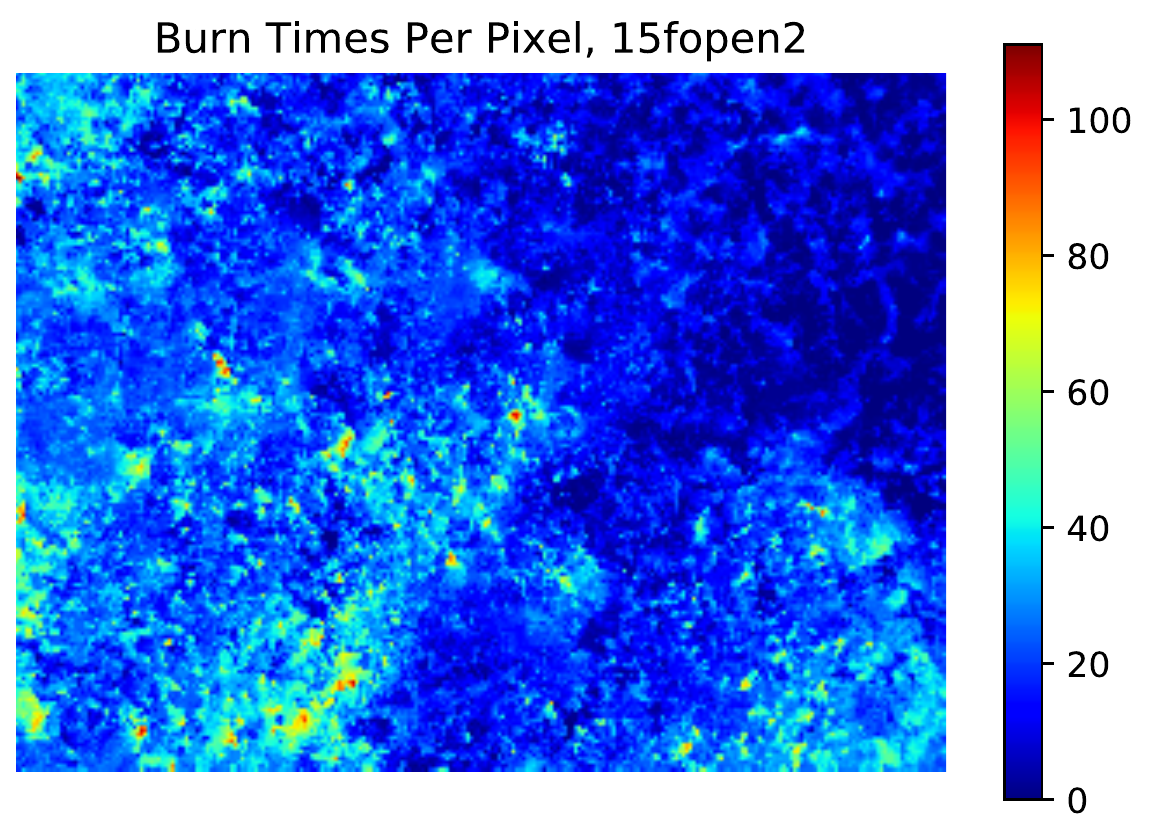}
\caption{\label{fig:burnTimes} The burn times for each pixel using a
500$^\circ$C threshold.  The colors indicate the number of time steps
that each pixel burned.}
\end{figure}

With a value separating burning and non-burning states, we can compute
the burn time of each pixel for the same burn.
Figure~\ref{fig:burnTimes} shows the number of time steps that each
pixel was above the 500\textdegree{}{C} threshold. Our estimates of burn
time are directly relevant to the fire spread rate since longer burning
time can overcome low probability of ignition and maintain fire spread.
For instance, Almeida and Macau's stochastic model~\cite{alm-mac2011}
contains a universal parameter for the probability that a pixel burns
(their ``B"), which is a representation of the burn time, and this
exerts an important control on fire spread in their model.

\begin{figure}
  \centering
  \begin{tabular}{cc}
  \includegraphics[width=0.5\textwidth]{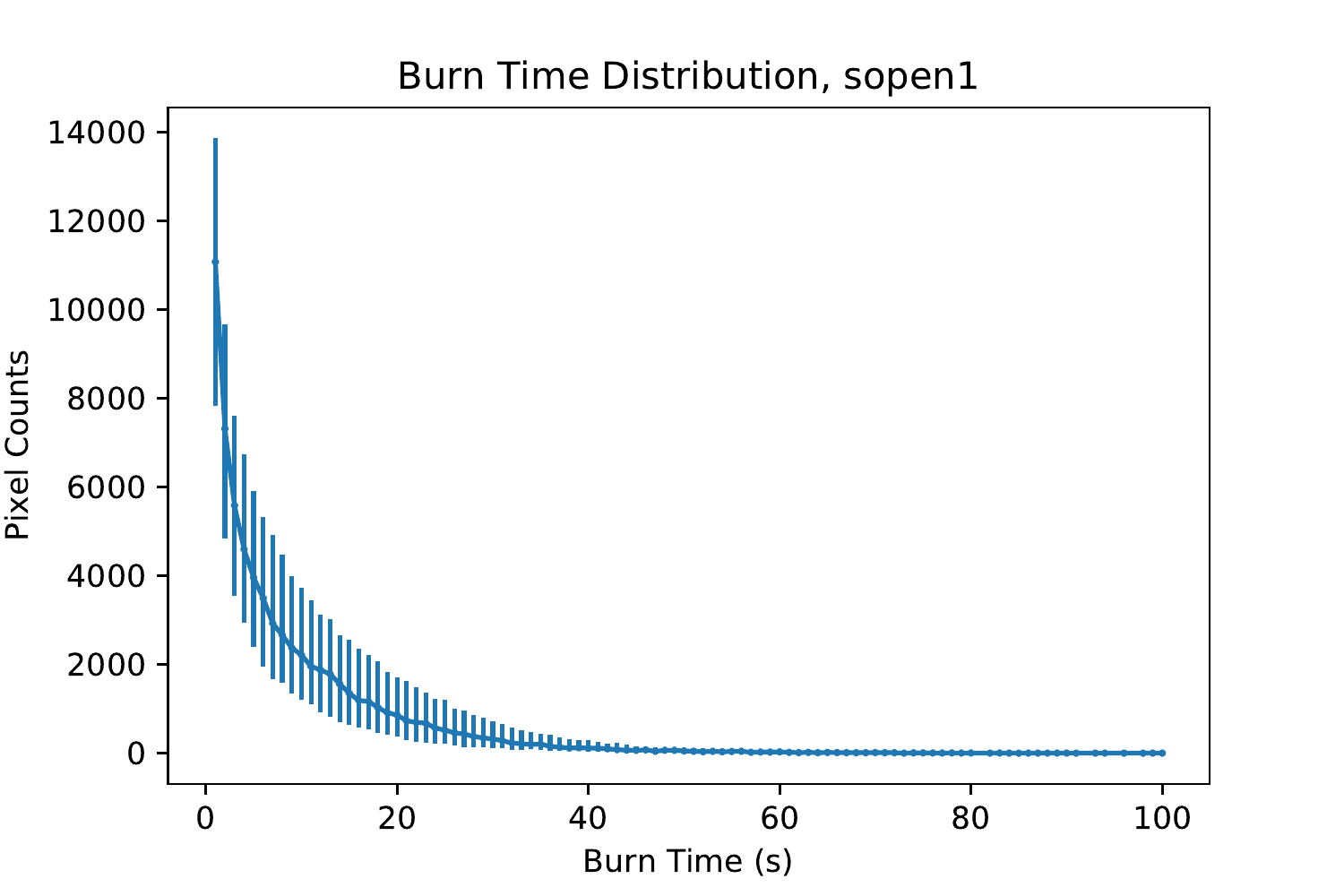} &
  \includegraphics[width=0.5\textwidth]{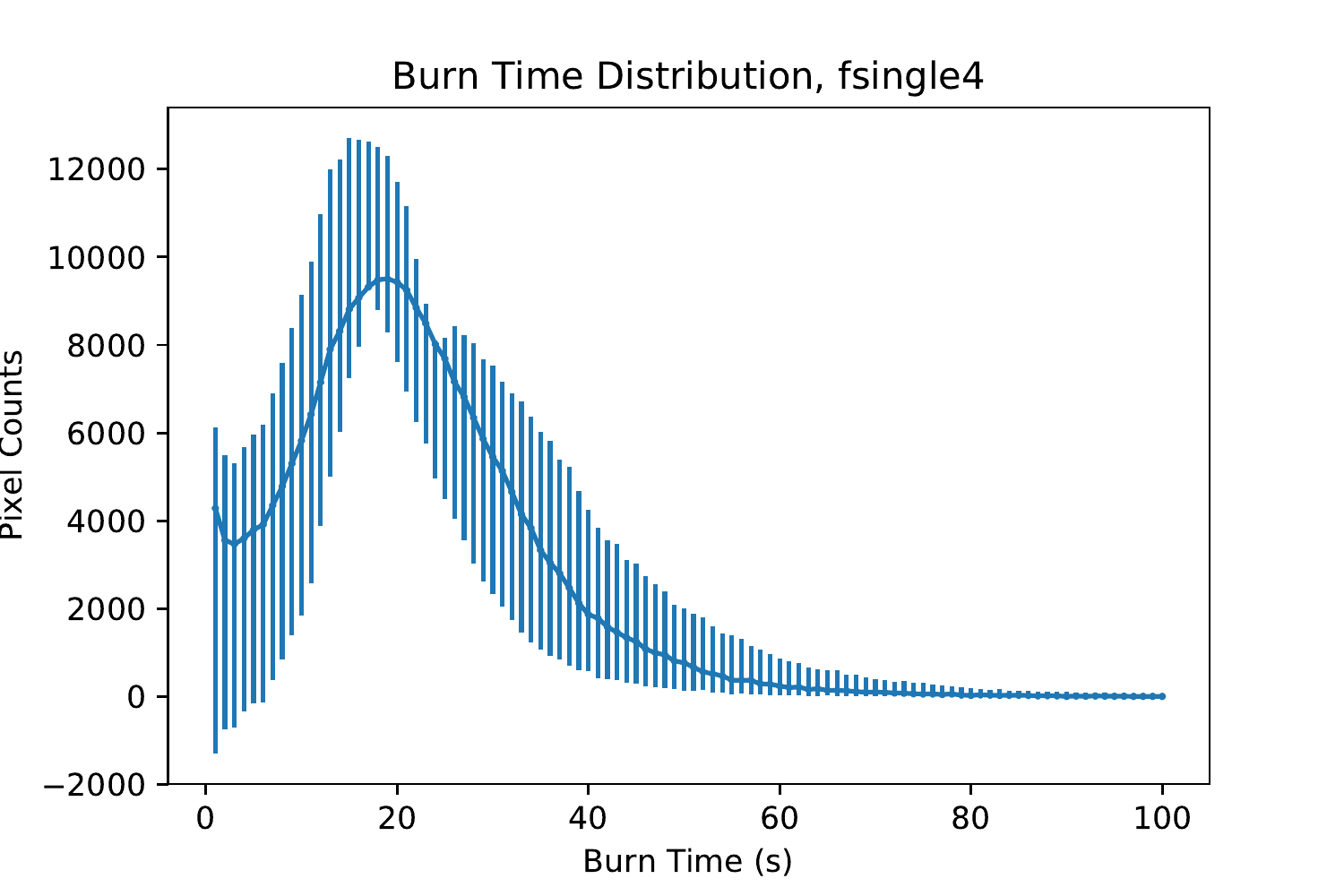} \\
  \includegraphics[width=0.5\textwidth]{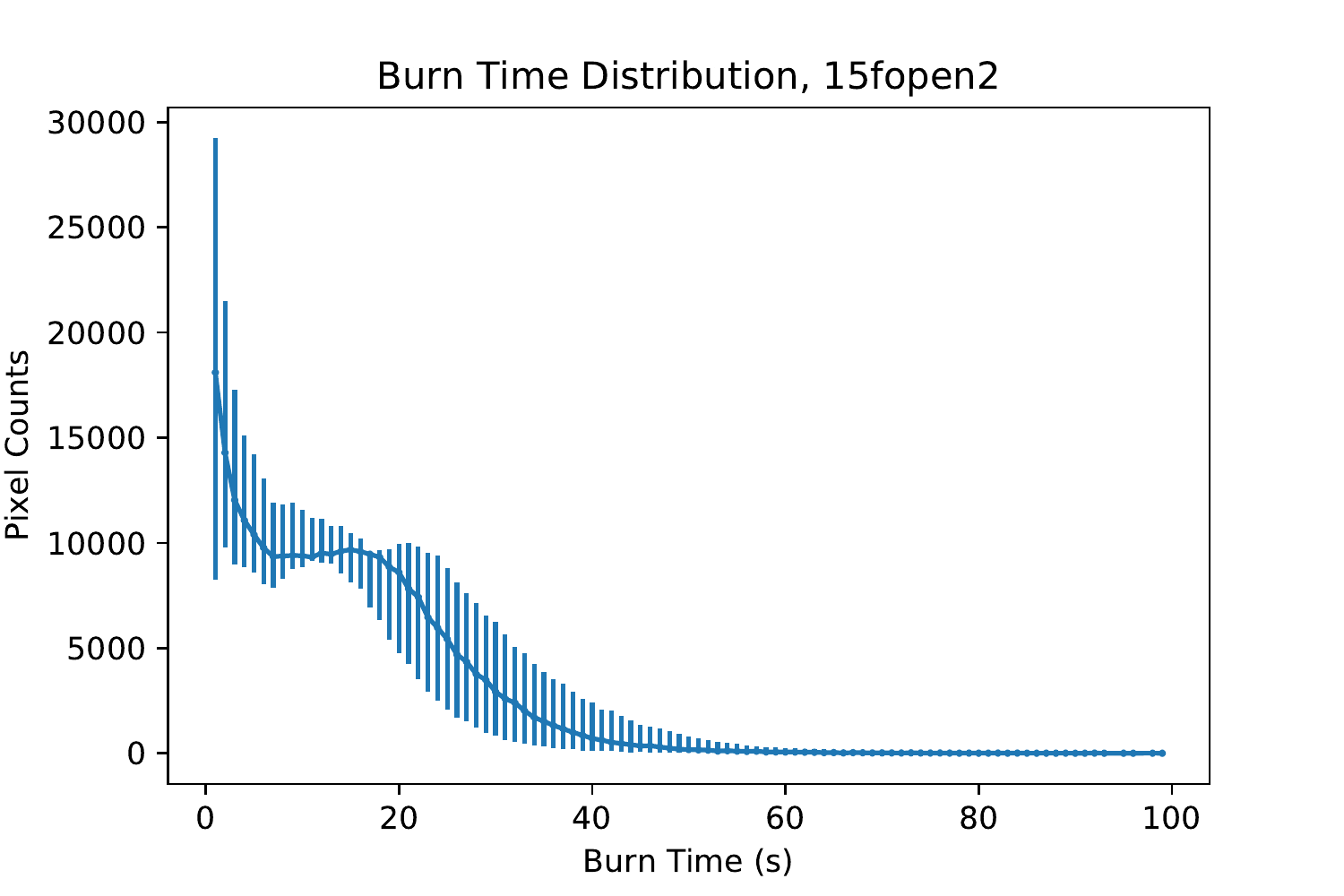} &
  \includegraphics[width=0.5\textwidth]{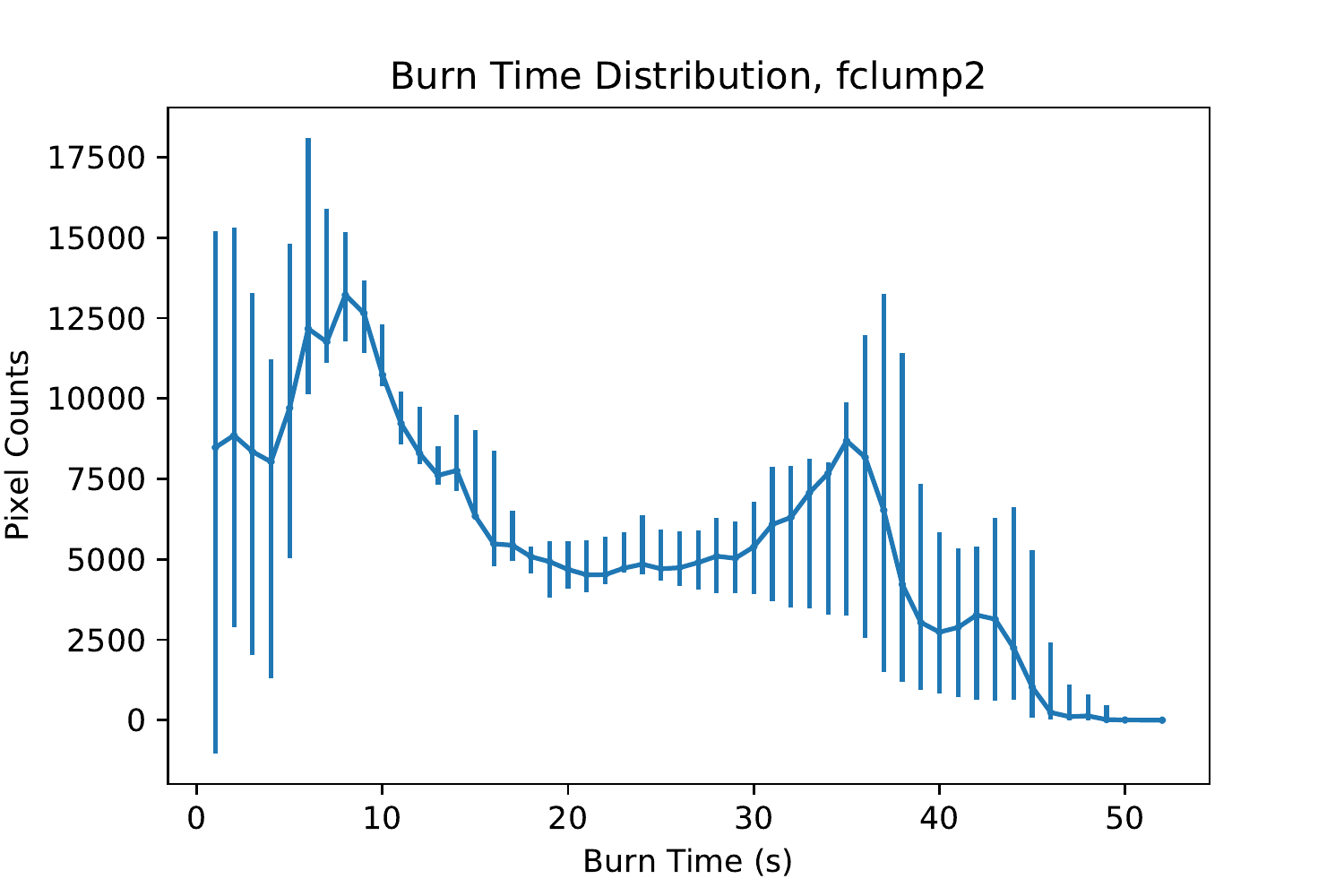}
  \end{tabular}
  \caption{\label{fig:burnDistributions} Four burn time distributions
    with $50^{\circ}$C error bars that are representatives of the 40
    datasets.  {\em Top left:} A burn time distribution with exponential
    decay with an $e$-folding time of approximately four seconds.  This
    is the most common distribution and can be found in twenty-four of
    our datasets.  {\em Top Right:} Burn time distribution with a
    characteristic single peak in the distribution.  This behavior is
    found in six of our datasets. {\em Bottom Left:} Burn time
    distribution with the characteristic of leveling off for several
    seconds before decreasing roughly exponentially. This behavior is
    found in eight of our datasets, including the dataset {\em 15fopen2}
    in Figures~\ref{fig:basicStats} and~\ref{fig:burnTimes}.  {\em
    Bottom Right:} Burn time distribution with the bi-modal
    characteristic of two distinct peaks in the burn time distribution.
    Only one fire shows this characteristic.}
\end{figure}

In addition to spatial maps of pixel burn times, we present burn time
distributions of the 40 datasets.  These distributions can be organized
into four general trends, and representative distributions from each
group are shown (Figure~\ref{fig:burnDistributions}).  The error is
estimated with a $\pm50^{\circ}$C range on the defined burn threshold.
Note that the maximum burn time in the bottom right distribution is
about half the size of the other three distributions.  These
distributions are described further Section~\ref{sec:discussion}.

\subsection{Probability of Spread}
Another key parameter of a CA model is the probability of spread.  We
compute probabilities of a pixel igniting when any number of its nearest
eight neighbors are on fire.  Spotting can be included by defining the
nearest neighbors to include additional layers of nearby pixels.  In
this work, however, we will coarsen the grid by aggregating pixels
(Section~\ref{sec:aggregation}), and this has the effect of reducing the
amount of spotting.

\begin{figure}
\centering
\begin{tabular}{cc}
  \includegraphics[width=0.5\textwidth]{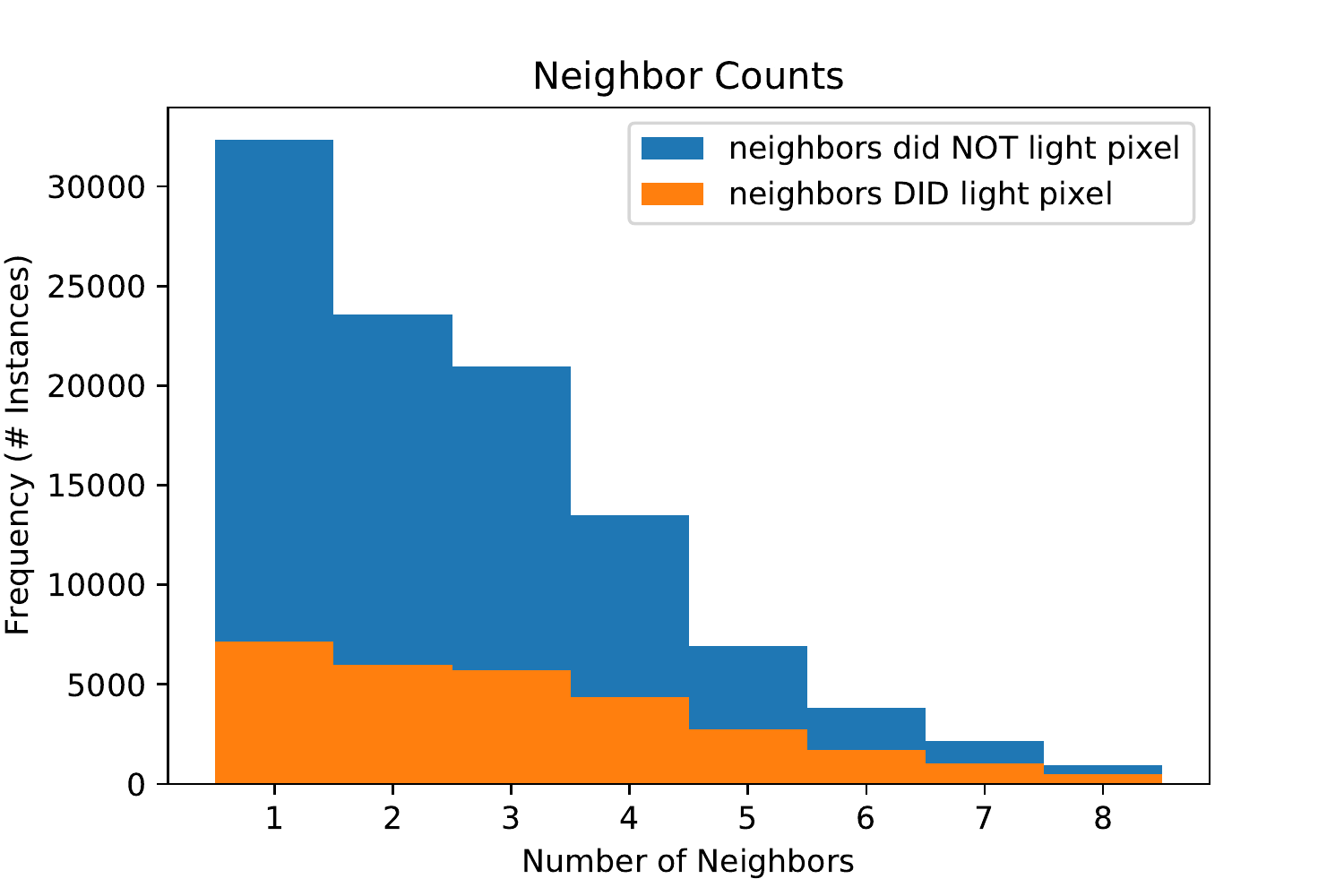} &
  \includegraphics{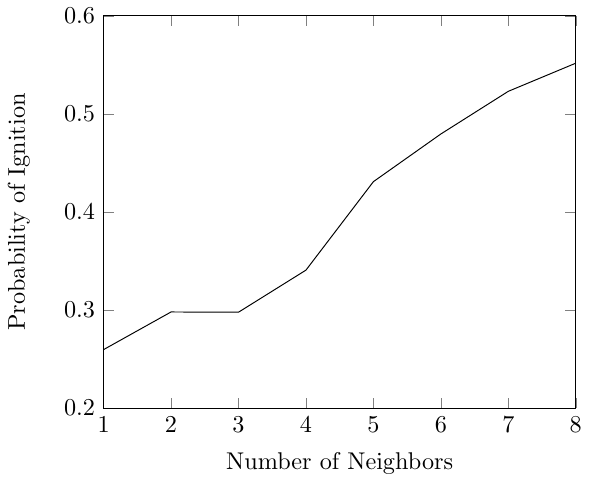}
\end{tabular}
\caption{\label{fig:spread1} {\em Left:} The number of instances where
  an unburnt pixel's nearest neighbors did (orange) and did not (blue)
  cause ignition.  {\em Right:} The probability of a pixel igniting
  conditioned on the number of its neighbors being on fire.  For
  example, if four of an unburnt pixel's neighbors are on fire, then it
  has approximately a 35\% chance of igniting at the next time step.}
\end{figure}

To calculate the probability of a fire spreading between pixels, we
first identify pixels that are newly ignited in each frame. For each of
these pixels, we count the number of nearest neighbors that were on
fire.  This gives a statistic of the likelihood of a pixel igniting
given the condition that some number of its neighbors are burning.  We
can perform the same count for pixels that did not ignite when at least
one of its neighbors was on fire.  We plot these two values as a
histogram in the left side of Figure~\ref{fig:spread1}.  We exclude the
case of newly ignited pixels where zero neighbors were on fire since
this corresponds to spotting, a phenomenon that we are not investigating
in this analysis.  In the right plot of Figure~\ref{fig:spread1}, the
histogram is used to compute the distribution of a pixel igniting.

\section{A Simple Model for Fire Spread}
\label{sec:simpleModel}
A factor in fire spread that we have not explicitly treated in this
analysis is wind.  Wind can be modelled in a CA code by choosing
probabilities of one pixel igniting another that are directionally
dependent.  For example, if there is no wind, then every burning
neighbor of an unburnt pixel has the same probability of igniting this
pixel.  However, if there is a strong eastward wind then the burning
neighbors to the left are the most likely to ignite an unburnt pixel.

We propose a basic two parameter model to understand the effect of the
wind on the data.  The parameters are the following.
\begin{enumerate}
  \item $M$ is the number of neighbors that have the ability of igniting
    an unburnt pixel.
  \item $p$ is the probability that one of these neighbors ignites an
    unburnt pixel.
\end{enumerate}
The two extreme cases are $M=1$, where only one neighbor can ignite a
pixel (high wind case), and $M=8$ where any pixel can ignite a
neighboring pixel (no wind case).  We let $X$ be the event of an unburnt
pixel igniting at the next time step and $Y$ be the number of neighbors
of this unburnt pixel that are currently burning.  For a fixed $p$ and
$M$, we are interested in computing $p(X \:|\: Y = N)$ which is the
probability that an unburnt pixel ignites, given that it currently has
$N$ burning neighbors.

These probabilities can be computed analytically.  For example, if only
one neighbor can ignite a pixel ($M=1$), then
\begin{align*} 
  p(X \:|\: Y = N) = \frac{N}{8}p.
\end{align*}
In the other extreme, if any neighbor can ignite a pixel ($M=8$), then
\begin{align*} 
  p(X \:|\: Y = N) &= \left(1 + (1-p) + \cdots + (1-p)^{N-1} \right) p \\
  &= \frac{1-(1-p)^N}{1- (1-p)} p \\
  &= 1 - (1-p)^N.
\end{align*}
The expressions for these probabilities become unwieldy for other values
of $M$, so we use a Monte-Carlo approach to compute the probabilities.
For a fixed $M$, $p$, and $N$, $10^6$ randomly selected configurations
of burning neighbors are chosen, and for each burning neighbor that is
able to ignite the unburnt pixel, we draw a random number to determine
if the unburnt pixel is ignited.  This method is verified with analytic
expressions for the probabilities with $M=2$ and $M=3$ (not reported in
this paper).

To find the best possible probability for each value of $M$, we compute
the value of $p$ that minimizes the root mean square (RMS) for $N =
1,\ldots,8$ between the Monte-Carlo method and the experimental data in
Figure~\ref{fig:spread1}.  This is done by choosing the optimal $p$ at
an equispaced partitioning of $[0,1]$, and then using the value of $p$
that minimizes the RMS to choose lower and upper limits for a new search
region.  We iterate this procedure until the first two digits of the RMS
is unchanging.  For each value of $M$, we report the best probability of
spread $p$ and the RMS value in Table~\ref{tbl:prob1}.  In
Figure~\ref{fig:prob1_model}, we plot the experimental data and the best
model fits when $M=1,2,7$.  The other values of $M$ are similar to
$M=7$.  While the smallest RMS value is achieved when seven neighbors
can ignite an unburnt pixel, this RMS value is comparable to other
simulations where fewer neighbors have the ability to ignite a pixel.
In particular, even if only 3 pixels can ignite an unburnt pixel, a RMS
of almost 20\% can be achieved for experimental data from complex
dynamics with a very simple CA model.  Further discussion of this
behavior is described in Section~\ref{sec:discussion}.

\begin{table}
\begin{minipage}{0.48\textwidth}
  \centering
  \begin{tabular}{|ccc|}
    \hline
    $M$ & $p$ & RMS \\
    \hline
    1 & 0.6394 & 0.2563 \\
    2 & 0.3772 & 0.2139 \\
    3 & 0.2671 & 0.2016 \\
    4 & 0.2059 & 0.1952 \\
    5 & 0.1679 & 0.1928 \\
    6 & 0.1402 & 0.1897 \\
    7 & 0.1192 & 0.1885 \\
    8 & 0.1100 & 0.1891 \\
    \hline
  \end{tabular}
  \caption{\label{tbl:prob1}The best probability of spread $p$ for each
  choice of the number $M$ of neighbors that can ignite an unburnt
  pixel.  Also reported is the root mean square error.}
\end{minipage}
\hfill
\begin{minipage}{0.5\textwidth}
  \centering
  \includegraphics{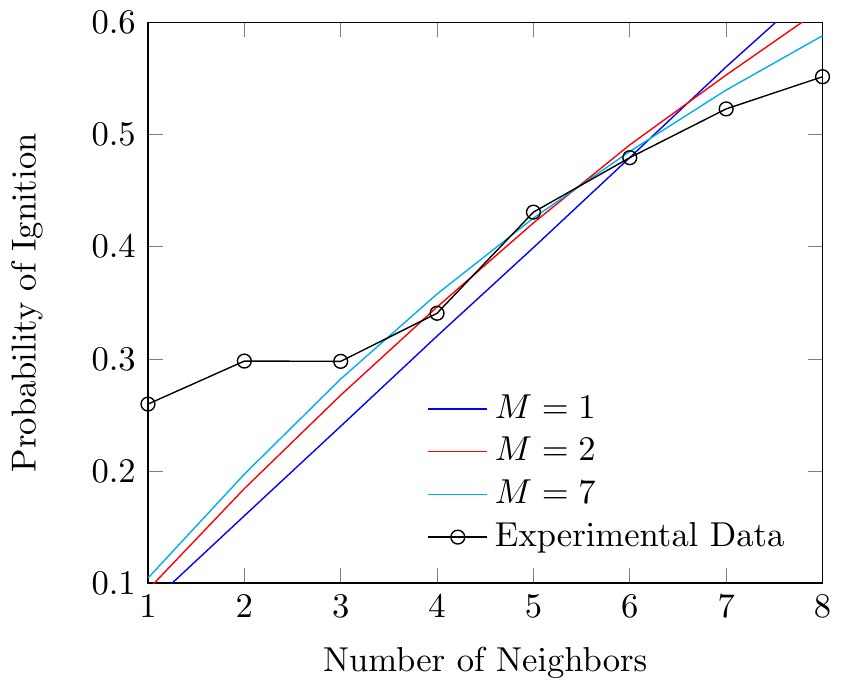} 
  \captionof{figure}{\label{fig:prob1_model}The probabilities of
  ignition for a different number of neighbors that can ignite an
unburnt pixel (solid lines).  The circled line is the experimental data
from Figure~\ref{fig:spread1}.} 
\end{minipage}
\end{table}

\section{Pixel Aggregation}
\label{sec:aggregation}
By only using the eight nearest neighbors as potential ignition sources
for an unburnt pixel, we are ignoring ignition due to spotting.  Given
the spatial resolution of 1cm$^2$ of the IR images, it is unreasonable
to assume that fire spreads to only its neighboring pixels at the 1 Hz
speed of the camera.  In reality, spotting occurs due to radiative and
convective transport of heat, or by transport of an ember.  To
investigate the effect of spotting and to examine the fire dynamics at
larger spatial scales, we aggregate groups of $3 \times 3$, $5 \times
5$, and $10 \times 10$ pixels.  We use the maximum temperature of the
pixels in the aggregate to assign a temperature to the aggregate.  The
maximum value is used instead of the mean since the mean would be skewed
by the 200\textdegree{}{C} threshold of the IR camera.  We repeat the
statistics reported in Figure~\ref{fig:spread1} for these new aggregated
pixels, and the results are in Figure~\ref{fig:spread2}.

We also repeat our probabilistic model for the aggregated pixel data
sets.  As before, for each number $M$ of pixels that can ignite an
unburnt neighbor, we find the probability $p$ that minimizes the RMS
over the number of burning neighbors $N$.  For the $10 \times 10$
aggregation, we do not include the $N=8$ entry in the RMS as this value
seems to be an outlier which is feasible since there are very few
instances when a $10 \times 10$ unburnt aggregated pixel has 8 burning
neighbors (see Figure~\ref{fig:spread2-10}).  The model fits are in
Figure~\ref{fig:probAggregated_model} and the optimal probabilities and
RMS are in Table~\ref{tbl:prob_model}.

\section{Discussion}
\label{sec:discussion}
Empirical data is needed to validate and inform new modelling tools in
wildland fire~\cite{cru-ale-sul2017}.  Cellular automata codes rely on
probabilistic rules to determine fire dynamics, and we have used 40 IR
datasets from controlled burns to analyze two of the main parameters
common in fire modeling using CA methods.  We have focused on the burn
times of a pixel and the probability of a unburnt pixel being ignited by
its burning neighbors.  We also analyzed the effect of the spatial
resolution on these probabilities.  Finally, we compared a simple model
for fire spread to the experimental data at different aggregate
resolutions. 

First, the burn times depend strongly on the fuel type, moisture, wind
speed, and other factors~\cite{koo2005}, but we found that from these 40
datasets, the burn times could be binned into four distributions plotted
in Figure~\ref{fig:burnDistributions}.  Most of the datasets' burn times
have a distribution with a small mean (only a few seconds) and
exponential decay (top left).  This corresponds to a fast burning
homogeneous fuel type.  Another set of fire observations are represented
by a mean closer to 20 seconds with an exponential decay (top right).
This corresponds to a slower burning homogeneous fuel type.  The third
distribution (bottom left) has both a slow burning and fast burning
fuels which indicates a heterogeneous fuel type.  Finally, a single data
set has a distribution with a large variance and a shorter maximum burn
time (bottom right), and this corresponds to an even more heterogeneous
fuel type.  The 40 datasets have several variables, including the
density of the trees and the moisture of the vegetation.  Most of the
datasets whose distribution resembles the top left plot of
Figure~\ref{fig:burnDistributions} come from the drier vegetation, while
the all of the datasets that resemble top right plot are from plots with
a wetter vegetation.

One conclusion that our analysis reveals is that a single parameter for
the burn time probability---many CA models make this assumption---is not
applicable since the burn time distributions are wide.  However, these
prior burn time distributions (Figure~\ref{fig:burnDistributions}) can
be improved by using the experimental data to create a posterior
distribution.  That is, by using a Bayesian approach, the burn times,
which are related to the fuel type, can be incorporated.

\begin{figure}
\centering
\begin{tabular}{cc}
  \begin{subfigure}[b]{0.5\textwidth}
  \centering
  \includegraphics[width=\textwidth]{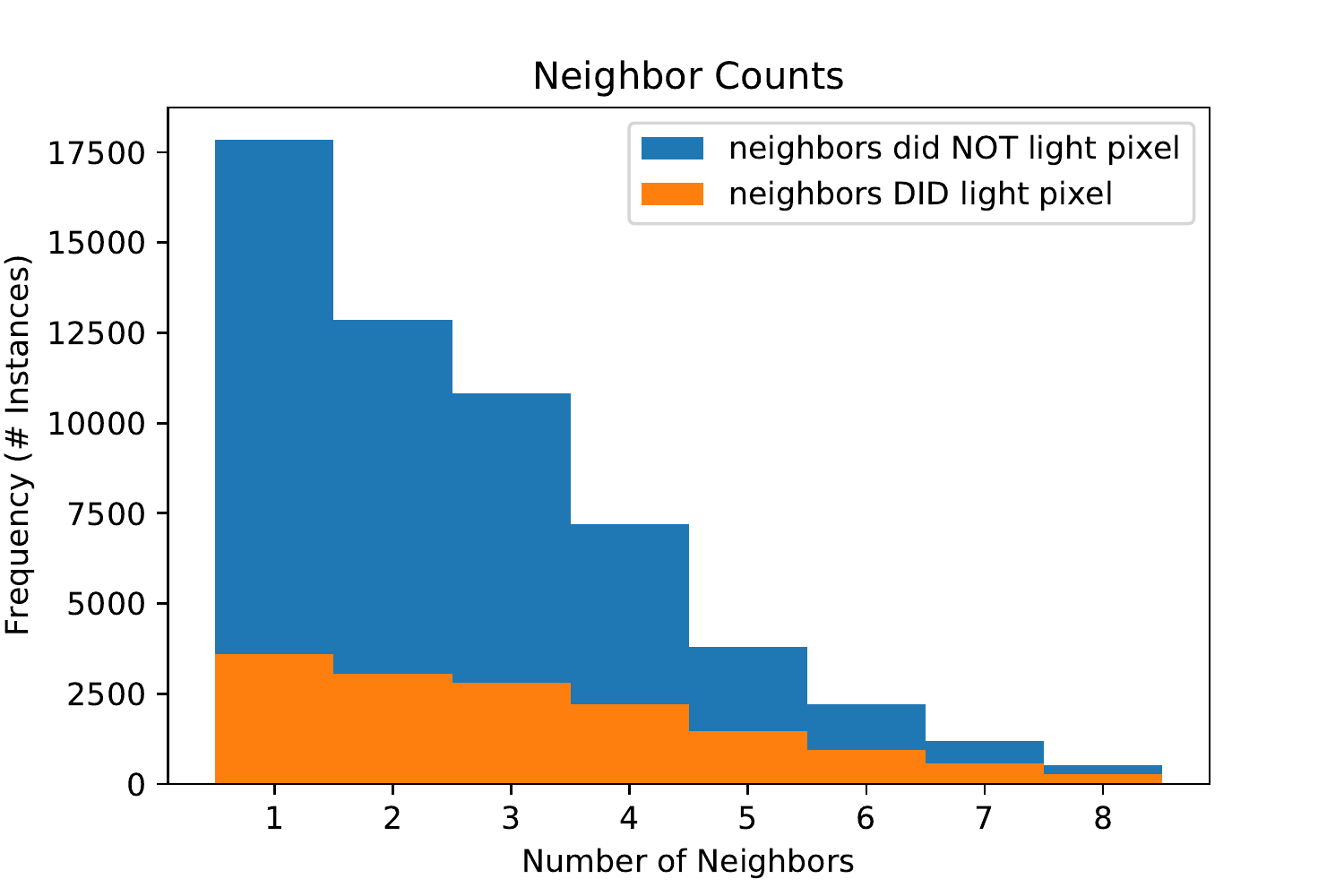}
  \caption{$3 \times 3$ aggregated groups.}
  \end{subfigure} 
  &
  \begin{subfigure}[b]{0.5\textwidth}
  \centering
  \includegraphics{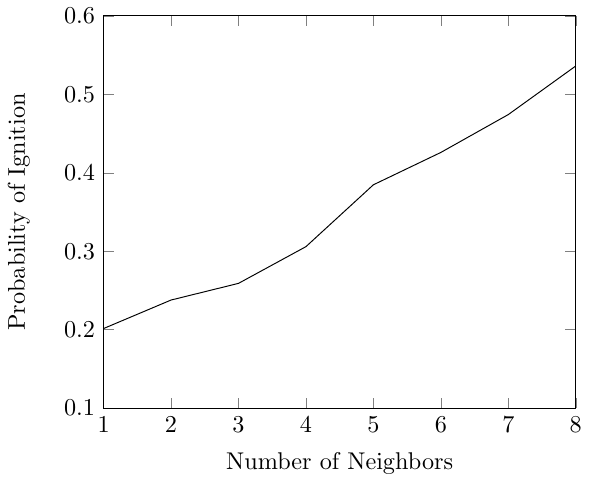}
  \caption{$3 \times 3$ aggregated groups.}
  \end{subfigure} \\

  \begin{subfigure}[b]{0.5\textwidth}
  \centering
  \includegraphics[width=\textwidth]{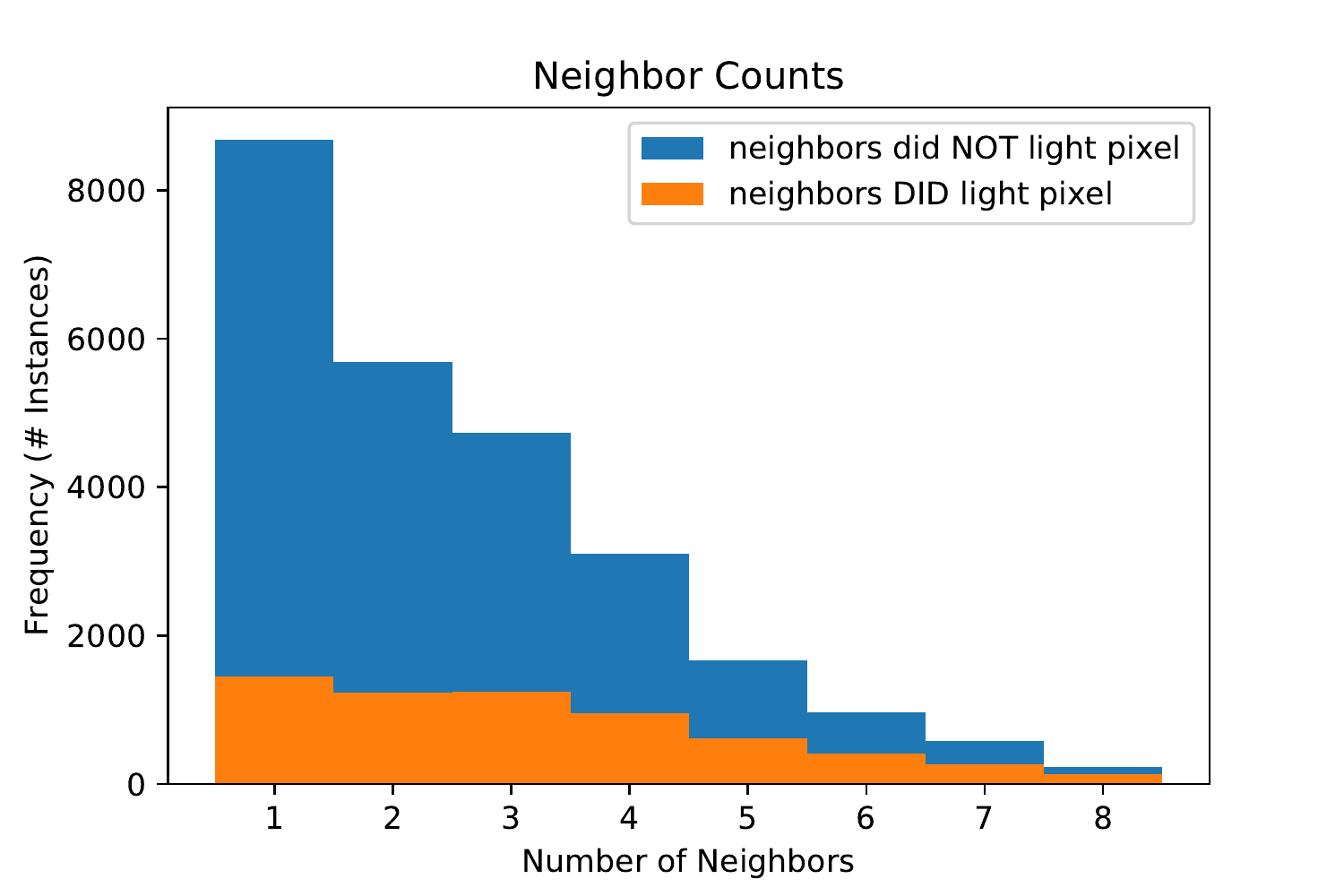}
  \caption{$5 \times 5$ aggregated groups.}
  \end{subfigure} 
  &
  \begin{subfigure}[b]{0.5\textwidth}
  \centering
  \includegraphics{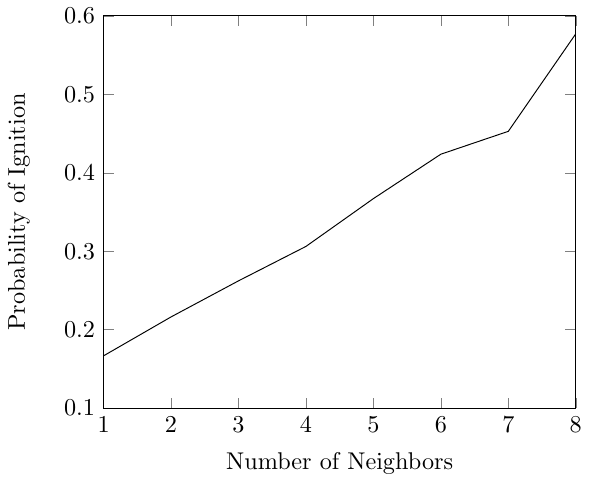}
  \caption{$5 \times 5$ aggregated groups.}
  \end{subfigure} \\

  \begin{subfigure}[b]{0.5\textwidth}
  \centering
  \includegraphics[width=\textwidth]{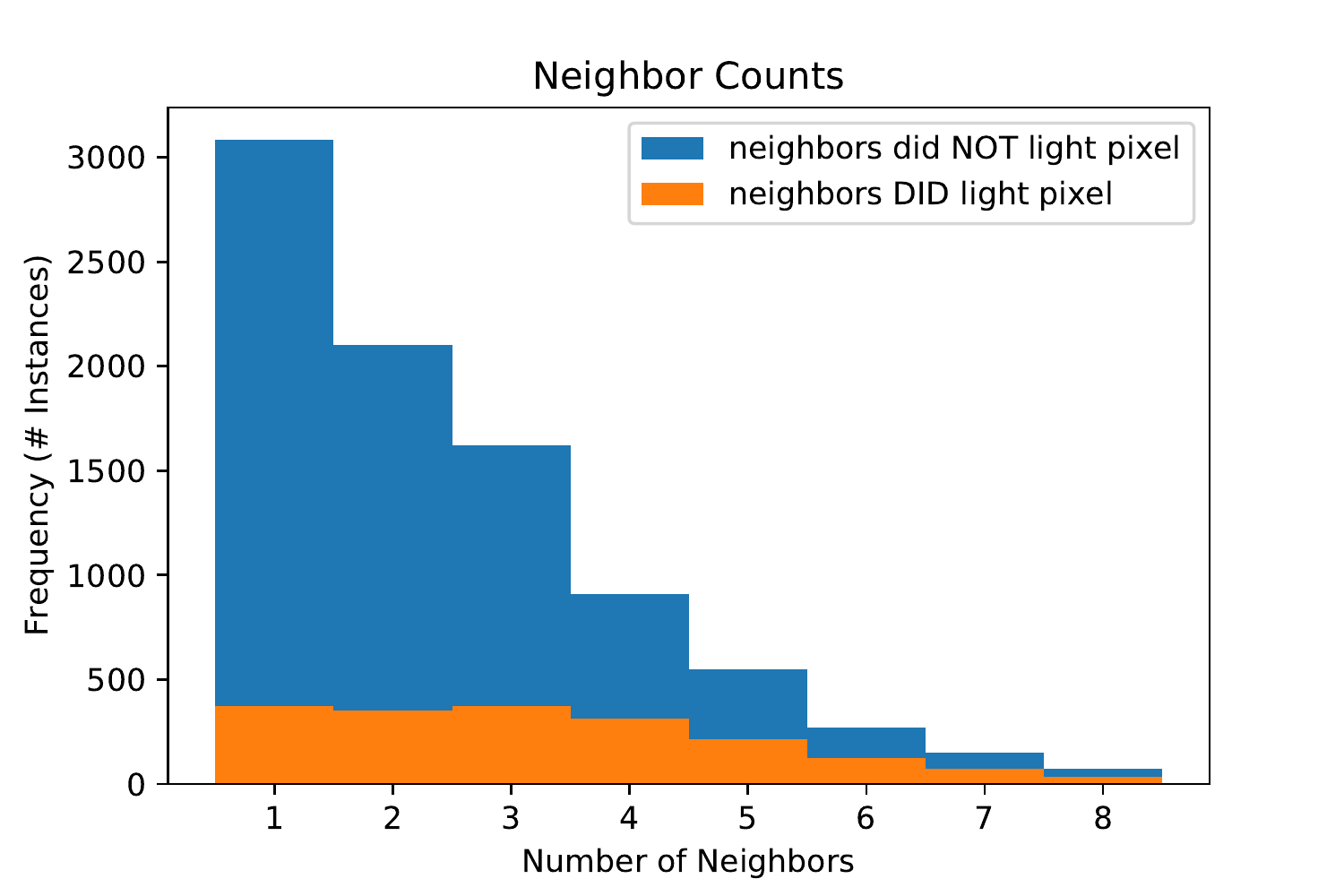}
  \caption{\label{fig:spread2-10} $10 \times 10$ aggregated groups.}
  \end{subfigure} 
  &
  \begin{subfigure}[b]{0.5\textwidth}
  \centering
  \includegraphics{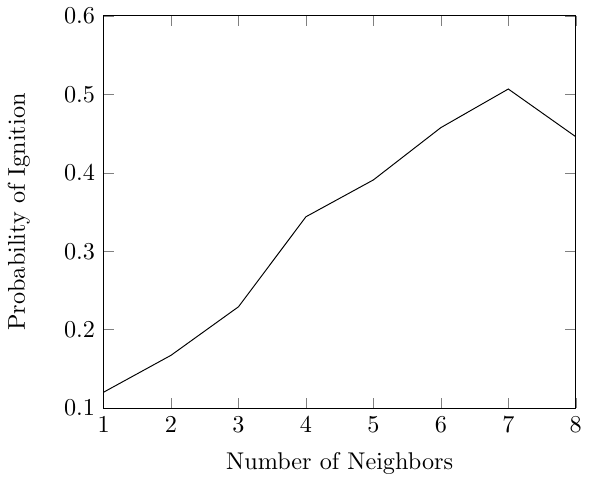}
  \caption{$10 \times 10$ aggregated groups.}
  \end{subfigure}
\end{tabular}
\caption{\label{fig:spread2} {\em Left:} The number of instances where
  an unburnt aggregated pixel's nearest neighbors did (orange) and did
  not (blue) cause ignition at different levels of aggregation.  {\em
Right:} The probability of an aggregated pixel igniting conditioned on
the number of its aggregated neighbors being on fire.}
\end{figure}

The probability of spread is described in terms of the number of burning
neighbors.  That is, given a newly ignited pixel, we determine how many
of its neighbors are burning at the previous time step.  The most common
occurring instance is only one burning neighbor, while eight burning
neighbors happens less frequently.  To analyze these probabilities for
larger grids, we aggregated pixels and computed the same probabilities.
At all the resolutions we observed that most pixels are ignited by only
one burning neighbor, while the fewest are ignited by eight neighbors.
Moreover, the probability of ignition (right plots of
Figures~\ref{fig:spread1} and~\ref{fig:spread2}) verify, unsurprisingly,
that the probability of ignition increases with the number of burning
neighbors, and this probability distribution is essential for a CA
model.

More interestingly, a simple model for fire spread captures features of
these transition probabilities.  First, for the unaggregated pixels
(Figure~\ref{fig:prob1_model}), the model does a good job of capturing
the experimental data when there are three or more burning neighbors.
However, the model severely underestimates the probability of spread if
only one or two neighbors are burning.  In this case, we expect that
much of the fire spread is nonlocal from convective heating and
small-scale spotting, and our model does not account for these nonlocal
effects.  This hypothesis is further supported by analyzing the
aggregated cells.  In these cases, the effective spotting becomes less
prevalent, and we expect that the leading cause of new ignition is
neighboring pixels.

In Figure~\ref{fig:probAggregated_model} and Table~\ref{tbl:prob_model},
we see that the fit between the model and the experimental data improves
as larger pixels are considered.  In particular, an RMS of less than 5\%
is achieved when considering pixels that are 10cm$^2$.  This suggests
that the resolution of the camera was high enough to observe the fine
scale processes of combustion and that some averaging over these
processes is needed to produce stable statistics of fire spread, a
result that was confirmed by Loudermilk et
al.~\cite{lou-ach-obr-hie-hor2014}. The degree of averaging necessary
was found to be 10cm$^2$ in these experiments.

The spatial structure of burn times (e.g. Figure~\ref{fig:burnTimes}) is
a small-scale example of the fuel bed complexity that can result from
even the simplest of rules.  For instance~\cite{shenk2000} describes a
simulated tree population model with instantaneous fire spread and zero
burn time.  Real wildland systems have finite spread rate but their
model shows how the fuel itself (they simulated trees but it could be
any fuel) can spontaneously develop complex spatial structure.

Differentiating between the three prevalent types of fire (i.e.~head
fire, backing fire, and flanking fire) is important because the
different types of fire have different probabilities for spreading.
With a similar statistical analysis we can differentiate these different
types of fire spread with the appropriate images, we will be able to
further constrain these probabilities and use them to build models with
more intrinsic fidelity to fire propagation, and fewer unknown
parameters, leading to more accurate predictions.

\begin{table}
\begin{minipage}{0.48\textwidth}
  \begin{subtable}[t]{0.5\textwidth}
  \centering
  \begin{tabular}{|ccc|}
    \hline
    $M$ & $p$ & RMS \\
    \hline
    1 & 0.5779 & 0.1734 \\
    2 & 0.3321 & 0.1459 \\
    3 & 0.2339 & 0.1355 \\
    4 & 0.1810 & 0.1333 \\
    5 & 0.1436 & 0.1301 \\
    6 & 0.1247 & 0.1296 \\
    7 & 0.1061 & 0.1284 \\
    8 & 0.0948 & 0.1287 \\
    \hline
  \end{tabular}
  \caption{\label{tbl:prob3}$3 \times 3$ aggregated groups.}
  \end{subtable}

  \begin{subtable}[t]{0.5\textwidth}
  \centering
  \begin{tabular}{|ccc|}
    \hline
    $M$ & $p$ & RMS \\
    \hline
    1 & 0.5843 & 0.1389 \\
    2 & 0.3256 & 0.1111 \\
    3 & 0.2306 & 0.1052 \\
    4 & 0.1765 & 0.1005 \\
    5 & 0.1378 & 0.1001 \\
    6 & 0.1188 & 0.0956 \\
    7 & 0.1028 & 0.0959 \\
    8 & 0.0896 & 0.0938 \\
    \hline
  \end{tabular}
  \caption{\label{tbl:prob5}$5 \times 5$ aggregated groups.}
  \end{subtable}

  \begin{subtable}[t]{0.5\textwidth}
  \centering
  \begin{tabular}{|ccc|}
    \hline
    $M$ & $p$ & RMS \\
    \hline
    1 & 0.6137 & 0.0643 \\
    2 & 0.3463 & 0.0443 \\
    3 & 0.2403 & 0.0426 \\
    4 & 0.1825 & 0.0432 \\
    5 & 0.1498 & 0.0406 \\
    6 & 0.1261 & 0.0436 \\
    7 & 0.1090 & 0.0436 \\
    8 & 0.0956 & 0.0450 \\
    \hline
  \end{tabular}
  \caption{\label{tbl:prob10}$10 \times 10$ aggregated groups.}
  \end{subtable}

\caption{\label{tbl:prob_model}The best probability of spread $p$ for
  each choice of the number $M$ of neighbors that can ignite an unburnt
  aggregated pixel.  Also reported is the root mean square error.}

\end{minipage}
\hfill
\begin{minipage}{0.5\textwidth}
  \begin{subfigure}[b]{0.5\textwidth}
  \centering
  \includegraphics{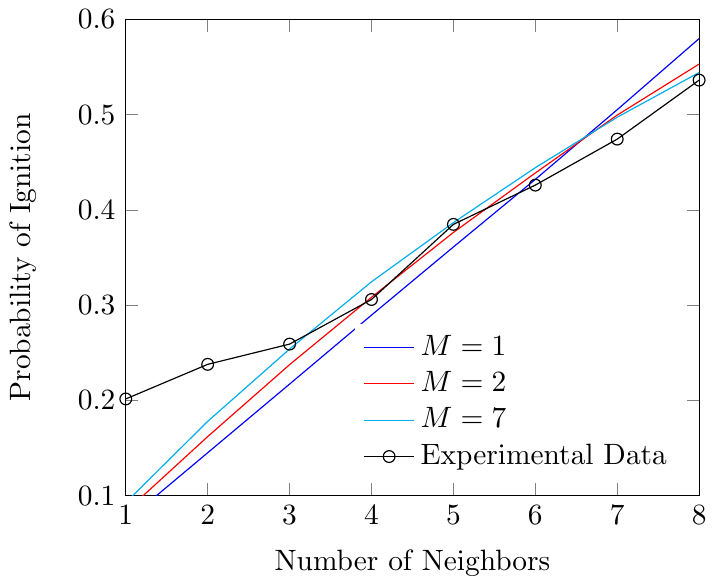}
  \caption{$3 \times 3$ aggregated groups.}
  \end{subfigure}

  \begin{subfigure}[b]{0.5\textwidth}
  \centering
  \includegraphics{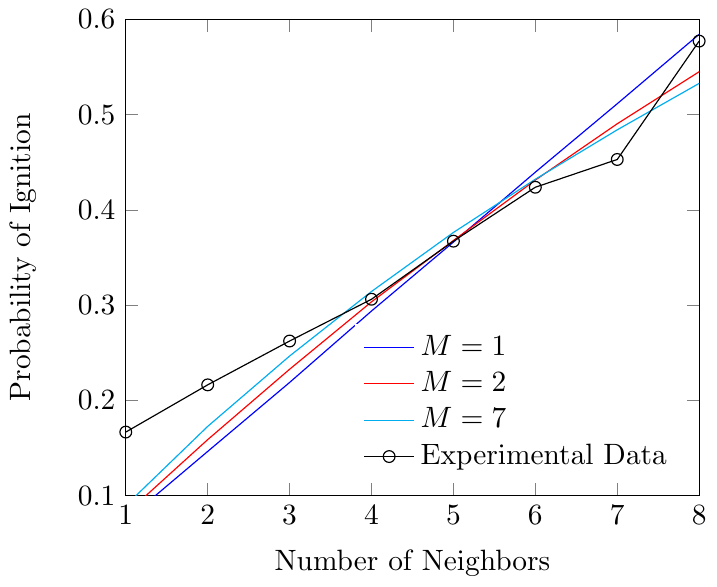}
  \caption{$5 \times 5$ aggregated groups.}
  \end{subfigure}

  \begin{subfigure}[b]{0.5\textwidth}
  \centering
  \includegraphics{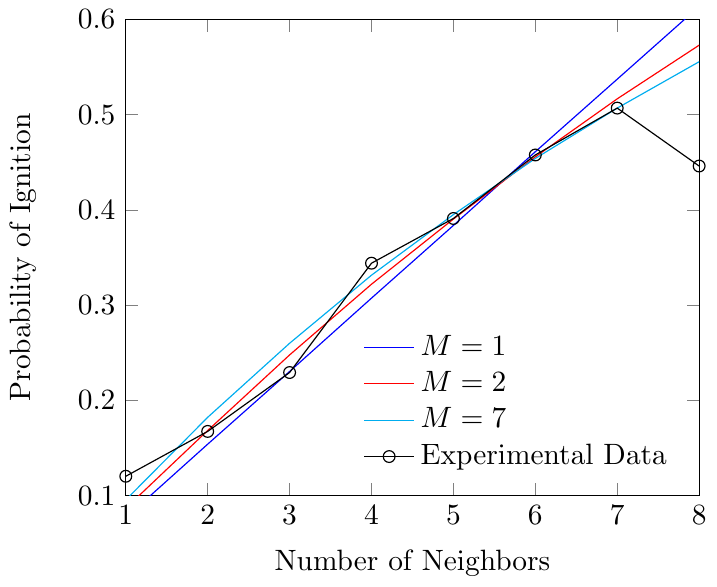}
  \caption{$10 \times 10$ aggregated groups.}
  \end{subfigure}
  \captionof{figure}{\label{fig:probAggregated_model}The probabilities
  of ignition for a different number of neighbors that can ignite an
unburnt pixel (solid lines).  The circled line is the experimental data
from Figure~\ref{fig:spread2}.} 
\end{minipage}
\end{table}

\biboptions{numbers,sort&compress}
\bibliographystyle{plain}

\end{document}